\documentclass[conference]{IEEEtran}
\IEEEoverridecommandlockouts
\usepackage[T1]{fontenc}
\usepackage{cite}
\usepackage{amsmath,amssymb,amsfonts}
\usepackage{algorithmic}
\usepackage{graphicx}
\usepackage{textcomp}
\usepackage{xcolor}
\usepackage{multirow}
\usepackage[a4paper, total={184mm,239mm}]{geometry}
\usepackage{fancyhdr}
\fancypagestyle{icecsheader}{%
    \fancyhf{}
    \fancyhead[C]{%
        \fontsize{8}{9}\selectfont
        2026 IEEE 33rd International Conference on Electronics, Circuits and Systems (ICECS)}

}
\fancypagestyle{icecsfirstpage}{%
    \fancyhf{}
    \fancyhead[C]{%
        \fontsize{8}{9}\selectfont
        2026 IEEE 33rd International Conference on Electronics, Circuits and Systems (ICECS)}
    \fancyfoot[L]{%
        \fontsize{8}{9}\selectfont
        979-8-3195-1905-4/26/\$31.00~\textcopyright{}~2026 IEEE}

}
\def\BibTeX{{\rm B\kern-.05em{\sc i\kern-.025em b}\kern-.08em
    T\kern-.1667em\lower.7ex\hbox{E}\kern-.125emX}}
\usepackage{todonotes}
\renewcommand{\baselinestretch}{0.94}
\begin{document}

\title{EBL: Efficient Broad Learning for Distributed Adaptive Harmonic Analysis
}

\author{
Changhong Li,
Georgios Floros,
Biswajit Basu,
Shreejith Shanker \\ 
Reconfigurable Computing Systems Lab, Electronic \& Electrical Engineering\\
Trinity College Dublin, Ireland\\
Email: \{lic9, florosg, basub, shreejith.shanker\}@tcd.ie
\thanks{This work was supported by the Sustainable Energy Authority of Ireland under Grant number 24/RDD/1170.}
\vspace{-3mm}
}

\maketitle
\IEEEpubid{%
    \makebox[\columnwidth][l]{%
        \fontsize{8}{9}\selectfont
        979-8-3195-1905-4/26/\$31.00~\textcopyright{}~2026 IEEE}%
    \hspace{\columnsep}%
    \makebox[\columnwidth]{}}
\thispagestyle{icecsfirstpage}

\begin{abstract}


Renewable energy systems and electrified transport have found widespread adoption in recent years.  
The integration of these non-linear loads, dominated by electric vehicle (EV) charging, however, has introduced severe harmonic distortion into the power grid, impacting the efficiency and lifetime of substation equipment and switchgear in the distribution network.
Rapid and high-precision harmonic analysis has hence become a prerequisite for effective harmonic control at the source of injection.
This paper proposes an Efficient Broad Learning (EBL) framework for distributed adaptive harmonic estimation.
As a quantised FPGA acceleration framework for BLS-style harmonic estimation, it offers high-accuracy estimation with half-cycle input, reconfigurable flexibility enabled by the FPGA implementation, and ultra-low latency, achieving 17.4 $\times$ faster predictions than the nearest reported FPGA method.
For harmonic prediction across multi-scenario charging and discharging nodes, the online transfer learning based on a closed-form solution rather than backpropagation in EBL demonstrates rapid adaptability.
By exploiting bespoke quantisation and sparsity, the approach consumes 5.9\% of the LUTs on the Zynq Ultrascale+ ZU7EV FPGA, using $\approx$ 82\% of the LUTs required by the state-of-the-art FPGA-accelerated estimator.
\end{abstract}
\vspace{-2mm}

\begin{figure*}[t]
    \centering
    \includegraphics[width=0.99\textwidth]{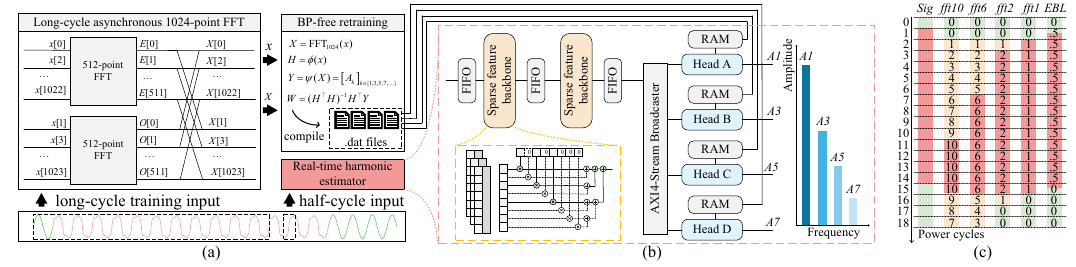}
    \caption{EBL overview: (a) Asynchronous long-window FFT for BP-free accelerator online adaptive retraining and half-cycle real-time harmonic estimation.
    (b) Dataflow accelerator with a sparse backbone and multiple reloadable task heads.
    (c) Latency cycles of harmonic estimation across different estimation methods.}
    \label{fig:framework} \vspace{-3mm}
\end{figure*}
\section{Introduction}\label{introduction}

The widespread adoption of electric vehicles (EVs) and distributed generation (solar energy, wind) have initiated an unprecedented integration of power and transportation systems~\cite{galus2008demand}. 
The growing penetration of EV charging facilities introduces dynamic nonlinearity into power grids, raising concerns over harmonic distortion in modern power systems. 
Harmonics, the integer multiples of its fundamental frequency, degrade power quality, accelerate equipment ageing, and threaten system stability and safety of the power grid. 
For power networks with significant EV charging integration, the resulting distortion can reduce the hosting capacity of the grid by up to 37.5\%~\cite{davila2023harmonic}. 
Accurate and timely harmonic estimation is a prerequisite for efficient harmonic mitigation and adaptive control in EV-integrated power systems.

Harmonic estimation involves analysing the distorted voltage/current signals to identify the harmonic components. 
IEC 61000-4-7 standard recommends the use of Discrete Fourier Transform (DFT) analysis as the reference framework for harmonic evaluation~\cite{iec200261000}.
In practice, the Fast Fourier Transform (FFT) remains a fundamental and effective method for harmonic analysis.
Advanced methods address leakage, noise, and non-stationarity through wavelet-based and Hilbert-transform-based analysis, or model-based techniques like ESPRIT~\cite{norman2012hybrid, gu2007estimating}.
These approaches typically require longer observation windows, higher computational cost, or stronger signal assumptions, hindering low-latency deployment in dynamic scenarios.

Several edge computing devices have been employed to offload conventional harmonic estimation approaches and improve their real-time performance.
The Discrete Wavelet Packet Transform (DWPT) \cite{tiwari2016hardware} and its variant Undecimated Wavelet Packet Transform (UWPT) \cite{tiwari2017fast} are offloaded to FPGAs for real-time harmonics estimation.
An FPGA-based DWPT implementation improves hardware efficiency through MAC-based wavelet filters and level-wise downsampling~\cite{baraskar2022digital}.
FPGA-based FFT analysis achieves a 33$\times$ speedup over an MCU implementation~\cite{li2024fast}.
Other algorithms, such as Gauss--Newton optimisation (GNO), have been implemented on DSPs for real-time harmonic estimation under power system frequency deviations~\cite{genccol2023efficient}.
However, these methods still require longer observation windows, limiting their application in dynamic harmonic estimation conditions prevalent in EV charging setups.


Parametric machine learning models have recently been explored for harmonic estimation and have been able to achieve higher accuracy than traditional methods with fewer sampling points.
Early neural-network methods such as radial basis function networks and adaptive wavelet neural networks improved estimation accuracy and reduced dependence on strong prior modelling assumptions~\cite{chang2009radial}. 
Support vector machine methods were also investigated for harmonic signal estimation because of their nonlinear modelling capability~\cite{katyara2020signal}. 
Graph-based models were also introduced to improve the estimation accuracy by capturing structural relationships in measurement data~\cite{madbhavi2023graph}.

Recently, Broad Learning System (BLS) based harmonic estimation has expanded the feature by broadening the network width, allowing feature extraction from shorter input windows without increasing network depth~\cite{li2023broad,chen2018broad}.
Its closed-form training enables fast model fitting.
However, these parametric machine learning models require higher computation and memory costs than conventional ML models, which leads to slower or more expensive hardware deployment.
In this paper, we propose EBL for efficient edge harmonic estimation.
Unlike earlier FPGA neural harmonic estimators~\cite{naoussi2009fpga,valtierra2013fpga}, this work focuses on a quantised FPGA implementation of a BLS-style estimator with a shared backbone and reloadable multi-task regression heads. We therefore make no claim of being the first neural-network-based edge harmonic estimator.
The main contributions of this paper are as follows:

\begin{itemize}
    \setlength{\itemsep}{0pt}
    \setlength{\parsep}{0pt}
    \item We propose EBL for efficient edge harmonic estimation, incorporating quantisation and sparsity for a hardware-efficient deployment without sacrificing performance.
    
    \item We implement EBL as a dataflow accelerator on an FPGA, which achieves ultra-fast, nanosecond-level estimation latency, making it ideally suited for embedding tightly coupled harmonic estimation and control at the source.
    
    \item We adapt the BLS-style fast closed-form solving and dynamic routing between the backbone network and task-specific heads, allowing the framework to adapt rapidly and cheaply to different harmonic estimation scenarios.
\end{itemize}
\vspace{-1mm}

\section{Methodology and Design} \label{methodology}

Fig.~\ref{fig:framework} summarises the proposed EBL framework for adaptive harmonic estimation.
EBL includes two components: a real-time harmonic estimation accelerator and a non-real-time adaptive learning component, as shown in Fig.~\ref{fig:framework} (a).
The real-time estimator is compressed through quantisation and unstructured sparsity, and then deployed as a dataflow accelerator shown in Fig.~\ref{fig:framework} (b) for hardware efficiency.
Broad feature backbone enables the estimation within half a power cycle, together with the low-latency inference, achieving high-accuracy estimation with faster response and tracking, as shown in Fig.~\ref{fig:framework} (c).
The frozen backbone is broadcast across multiple real-time regression heads with decoupled weights. 
These weights support online updates supervised by asynchronous long-window FFT, thereby enabling adaptive deployment across multiple scenarios. 

\subsection{Harmonic Estimation Model}

The model takes 32 samples (half-cycle) of the raw harmonic waveform as input for accurate estimation.
In contrast, conventional FFT-based methods require at least one cycle for acceptable estimation accuracy.
Fig.~\ref{fig:framework}(c) illustrates the temporal motivation of this design.
When the actual harmonic state changes, the raw signal reflects the transition immediately, while estimators relying on multi-cycle FFT accumulation respond only after several power cycles and therefore exhibit a delayed transition.
By contrast, the proposed EBL accelerator is driven by half-cycle inputs and a deeply pipelined dataflow engine, which allows the accelerator to generate estimates that are synchronised with the actual occurrence time of the harmonic event.
This makes the learned estimator suitable as the fast response path for edge control, while the longer-window FFT path remains responsible for slower but more reliable supervision and recalibration.

Inherited from the BLS, the backbone employs a broad feature layer and an enhancement layer to construct a high-dimensional feature space through extensive random mappings. 
The broad feature layer and enhancement layer are frozen for feature extraction, while the last head layers are updated with closed-form regression rather than backpropagation.S
BLS's feature enhancement relies on model width expansion, which can be implemented as a highly parallel and pipelined architecture.

Hybrid precision quantisation and unstructured sparsity are introduced to improve its deployment efficiency.
We apply 3-bit weight and activation quantisation to the target estimation model, while retaining 16-bit quantisation only for the output layer to preserve the estimation accuracy.
To further compress the model, we perform 85\% unstructured pruning.
These quantisation and pruning settings are determined via ablation studies to choose the most compressed configuration that incurs no accuracy loss relative to the FP32 model, yielding a theoretical compression ratio of 71$\times$ compared with the FP32 baseline.

\subsection{FPGA Dataflow Accelerator with Multi-head Reuse}


We use FINN~\cite{blott2018finn} to transform the generated sparse quantised model into a deployable streaming accelerator.
In the shared feature backbone, we adopt a fully unrolled weight-embedded Matrix-Vector-Activate-Unit (MVAU).
The sparse implementation adopts the unstructured sparsity method proposed in LogicSparse~\cite{li2025logicsparse}.
By embedding the sparsity into the network, the compiler can eliminate corresponding multipliers and adder tree logic to reduce the overall resource consumption and critical path. 
%
Beyond the hardware efficiency provided by unstructured sparsity, the output feature stream of the backbone is synchronously replicated through an AXI Stream Broadcaster.
The broadcaster allows multiple downstream regression heads to share the same features with reduced hardware overhead.

For the MVAU design of each regression task head, we implement a unified and reusable MVAU with decoupled weights and manually stitch these units after the broadcaster.
All task heads follow the same execution schedule, consume broadcast features at the same pipeline cadence, and maintain a fixed output interval to avoid blocking and pipeline stalls.
By configuring the weight mode as writable, the weights of each regression head can be dynamically loaded from LUTRAM and updated online by the control side during operation.

\subsection{Closed-form Online Adaptation and Dynamic Routing}

In FPGA-based quantised neural network accelerators, hardware generation typically requires synthesis, placement, and routing.
When the neural network is fine-tuned for a different task, the entire accelerator has to be regenerated with a highly time-consuming process.
However, this adaptation is necessary for harmonic estimation with varying operating conditions.

In accordance with the harmonic evaluation requirements of IEC 61000-4-7, we employ a longer waveform buffer and a non-real-time 1024-point FFT over 12 cycles on the PS control side to convert the harmonic signal to the frequency domain. This asynchronous 1024-point FFT teacher path is separate from the 32-sample half-cycle accelerator input path; the latter is used for real-time inference, whereas the former supplies slower supervision and recalibration.
The resulting frequency domain features are used to extract the amplitude and phase of each harmonic order, which serve as the training label vector.
The raw waveform is segmented into multiple samples according to the model input size, covering a half cycle, replicating the normal input format seen by the estimation model.
The backbone features are accumulated into a feature matrix, after which the task-head weights are updated through the closed-form solution of ridge regression, as shown in equation~\ref{eq:head_update}. 
Here, $\mathbf{H}$ denotes the stacked feature matrix and $\mathbf{Y}$ denotes the target matrix. 
Since only the task heads are updated through regression, online retraining becomes much more practical without resorting to costly backpropagation. The regularisation parameter $\lambda$ is selected on the validation split and is shared by the task heads.
\begin{equation}
\mathbf{W}_{\mathrm{head}}
=
\left(\mathbf{H}^{\top}\mathbf{H}+\lambda\mathbf{I}\right)^{-1}\mathbf{H}^{\top}\mathbf{Y},
\label{eq:head_update}
\end{equation}

After asynchronous PS retraining, the updated weights are compiled into \texttt{.dat} files and reloaded into the task-head LUTRAMs.
This supports fine-tuning across deployment scenarios and different regression tasks through software updates, without regenerating the accelerator.

\section{Experimental Results}\label{results}

\subsection{Experimental Setup}
We conduct experiments on two datasets to benchmark our EBL framework.
The first dataset is scenario A1 from BLS's evaluation as an ideal scenario for fair accuracy comparison with other approaches, especially BLS~\cite{li2023broad}. 
The fundamental frequency $f_0$ is set to 60\, Hz and allowed to vary within $\pm 0.5\%$, while the amplitudes of the first, third, fifth, and seventh harmonics fluctuate within $\pm 1\%$. Gaussian white noise with a signal-to-noise ratio (SNR) of 26\,dB is added.

Real-world current waveforms were collected using a PICO 4444 portable oscilloscope and an HIOKI 9660 probe, with each operating scenario forming a separate recording group for adaptation and validation.
The measurement results were obtained from an EV charger in the Grid to Vehicle (G2V) and V2G modes, and from a grid-connected battery in the Grid to Battery (G2B) and Battery to Grid (B2G) modes.
%

The model training and hardware compilation are conducted on a workstation equipped with NVIDIA RTX A4000 GPU and an Intel U7\textendash 265K CPU using Vivado~2022.2.
The backbone is frozen during fine-tuning, and only the regression-head weights are updated by the closed-form ridge-regression rule in (1).
The accelerator is generated targeting the AMD ZU7EV FPGA on the ZCU104 development kit.

\begin{figure}[t!]
    \centering
    \includegraphics[width=0.48\textwidth, trim={0 0.5mm 0 0.5mm}, clip]{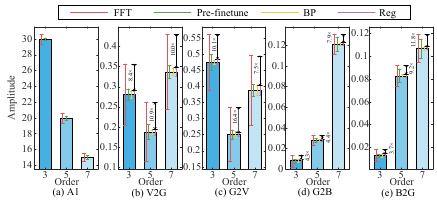}
    \caption{Estimated amplitudes of the 3rd, 5th, and 7th harmonics with 1-cycle FFT, EBL pre-trained from synthetic data (Pre-finetune), and EBL fine-tuned with closed-form regression under A1 (Reg), for different charging and discharging scenarios.}
    \label{fig:r1} \vspace{-3mm}
\end{figure}

\subsection{Synthetic Signal Harmonic Estimation}
The harmonic estimation relative error in the simulation scenario A1 is summarised in Table~\ref{tab:rel_error_methods}. For harmonic order $k$, the relative error is defined as $100|\hat{A}_k-A_k|/\max(|A_k|,\epsilon)$, where $A_k$ is the reference amplitude and $\epsilon$ prevents numerical instability for amplitudes close to zero. The entries in Table~\ref{tab:rel_error_methods} are percentages.
Performance of competing models are recreated from ~\cite{li2023broad}, and compared to the proposed EBL under the same input sequence for fair comparison. 
Despite 3-bit quantisation and 85\% unstructured pruning, the proposed compressed EBL model preserves accuracy close to the original BLS model~\cite{li2023broad}. 
It achieves the lowest mean relative errors on the 3rd and 5th harmonics, and the lowest maximum relative errors on the 3rd, 5th, and 7th harmonics. 
By contrast, FFT and WPT exhibit much larger errors, especially at higher harmonic orders. 
Competing methods use desktop-class processors and native precision, whereas EBL uses an edge-class ZYNQ device and mixed precision.

\begin{table}[t!]
    \centering
    \caption{Mean and maximum relative error (\%) of different harmonic estimation methods for each harmonic order}
    \label{tab:rel_error_methods}
    \footnotesize
    \setlength{\tabcolsep}{4pt}
    \begin{tabular}{lcccc|cccc}
        \hline
        \multirow{2}{*}{Method} & \multicolumn{4}{c|}{Mean relative error} & \multicolumn{4}{c}{Max relative error} \\
        \cline{2-9}
        & 1st & 3rd & 5th & 7th & 1st & 3rd & 5th & 7th \\
        \hline
        FFT   & 0.67 & 2.52 & 3.55 & 5.37 & 2.62 & 14.22 & 19.66 & 29.72 \\
        DWPT   & 0.67 & 2.57 & 3.84 & 4.62 & 2.62 & 14.51 & 20.41 & 33.36 \\
        MLP   & 0.83 & 0.80 & 0.82 & 0.79 & 2.00 & 2.00 & 2.02 & 2.00 \\
        RBF   & 0.53 & 0.50 & 0.56 & 0.48 & 1.24 & 1.09 & 1.29 & 1.10 \\
        AWN   & 0.50 & 0.50 & 0.53 & 0.48 & 1.02 & 1.03 & 1.09 & 1.01 \\
        BLS   & \textbf{0.49} & 0.50 & 0.52 & \textbf{0.48} & \textbf{1.00} & 1.01 & 1.01 & 1.00 \\
        \textbf{Prop.} & 0.51 & \textbf{0.48} & \textbf{0.51} & 0.50 & 1.09 & \textbf{0.97} & \textbf{1.00} & \textbf{0.99} \\
        \hline
    \end{tabular}
\end{table}

\subsection{Online Adaptation Across Operating Scenarios}

\begin{figure*}[t!]
    \centering
    \includegraphics[width=0.95\textwidth, trim={0 2.3mm 0 2.5mm}, clip]{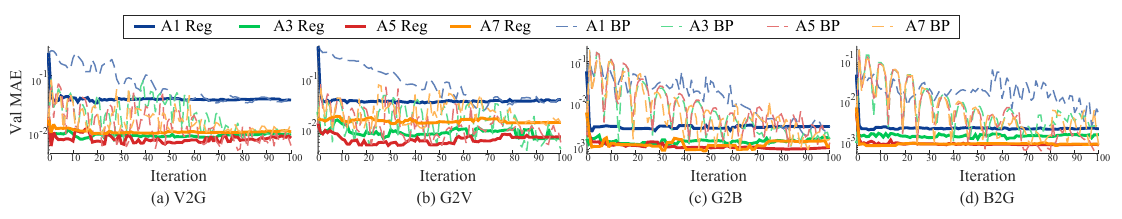}
    \caption{Validation MAE of model fine-tuning under different scenarios (A1 -- A7) using the proposed closed-form regression (Reg). Here A1--A7 denote operating scenarios, not harmonic amplitudes.}
    \label{fig:r2} \vspace{-0mm}
\end{figure*}

Fig.~\ref{fig:r1} presents the harmonic magnitude estimates obtained by different methods across all scenarios. 
The harmonic distributions vary from different scenarios, and the performance of the FFT degrades noticeably in the real high-noise scenarios. 
EBL can generalise effectively from the original simulation-trained setting to different real-world scenarios and achieves lower estimation errors than FFT.
After fine-tuning, its estimation error is further reduced.
This advantage is particularly evident in the G2V mode.
Taking the 5th-order harmonic as an example, the estimation error of EBL is reduced by $16.4\times$ compared with FFT. 
In most scenarios, the model updated through closed-form regression achieves better performance than the pre-trained model.
Fig.~\ref{fig:r2} shows the fine-tuning iterations under different scenarios.
The reported outputs evaluate harmonic amplitudes.
Closed-form regression fine-tuning requires fewer online resources because it updates only the task heads.

\subsection{Latency and Resource Efficiency}

Table~\ref{tab:latency_methods_devices} compares the latency results of our EBL with competing methods in the literature, where latency is reported. 
As shown, the sparse optimisation reduces the latency of the EBL implementation from 82\, ns (\textbf{Prop (d)}) to 70\, ns (\textbf{Prop (s)}). 
Both EBL implementations achieve lower latency than all reported harmonic estimation methods across different deployment devices, with \textbf{Prop (s)} achieving $\approx17.4\times$ improvement over the nearest FPGA-based harmonic estimation method in the literature,  DWPT~\cite{baraskar2022digital}. \vspace{-2mm}


\begin{table}[htbp]
    \centering
    \caption{Latency comparison of different methods and platforms}
    \label{tab:latency_methods_devices}
    \footnotesize
    \setlength{\tabcolsep}{2.2pt}
    \begin{tabular}{llr|llr}
        \hline
        Method & Platform & Latency & Method & Platform & Latency \\
        \hline
        GNO \cite{genccol2023efficient}  & F28027    & 0.31 ms          & FFT \cite{li2023broad}  & i5-9400F & 2.6 $\mu$s \\
        FFT \cite{li2024fast}            & ZYNQ7000 & 0.33 ms          & DWPT \cite{li2023broad} & i5-9400F & 10.4 ms \\
        DWPT \cite{tiwari2016hardware}            & AC-701     & 8.8 -- 13 $\mu$s          & MLP \cite{li2023broad}  & i5-9400F & 7 $\mu$s \\
        DWPT \cite{baraskar2022digital}  & AC-701    & 1.22 $\mu$s      & RBF \cite{li2023broad}  & i5-9400F & 22 $\mu$s \\
        \textbf{Prop. (d)}               & ZU7EV    & 82 ns            & AWN \cite{li2023broad}  & i5-9400F & 26.3 $\mu$s \\
        \textbf{Prop. (s)}               & ZU7EV    & \textbf{70 ns}   & BLS \cite{li2023broad}  & i5-9400F & 80.3 $\mu$s \\
        \hline
    \end{tabular}

\end{table}


The hardware resource comparison results are summarised in Table~\ref{tab:resource_utilization}. 
Unstructured sparsity reduces multiplier and adder-tree complexity, improving latency and reducing LUT and FF utilisation by 47\% and 44\%, respectively.
Compared with the FPGA baseline implementation based on DWPT for harmonic estimation~\cite{baraskar2022digital}, the sparse EBL achieves superior latency and accuracy while using fewer LUTs and DSPs in its implementation. \vspace{-2mm}

\begin{table}[htbp]
    \centering
    \caption{Resource utilisation of traditional approaches implemented on FPGAs, comparing with our EBL's dense and sparse versions.}
    \label{tab:resource_utilization}
    \footnotesize
    \setlength{\tabcolsep}{5pt}
    \begin{tabular}{lrrrrr}
        \hline
        Method & FF & LUT & LUT RAM & BRAM & DSP \\
        \hline
        DWPT \cite{tiwari2016hardware} & 38,655 & 35,208 & 178  & 2 & - \\
        UWPT \cite{tiwari2017fast} & \textbf{6,629} & 16,298 & - & 14 & 36 \\
        DWPT \cite{baraskar2022digital} & 11,007 & 15,019 & 1,325 & 18 & 52 \\
        \textbf{Prop. (d)}              & 24,445 & 22,901 & \textbf{132} & \textbf{0} & \textbf{8} \\
        \textbf{Prop. (s)}              & 13,711 & \textbf{12,249} & \textbf{132} & \textbf{0} & \textbf{8} \\
        \hline
    \end{tabular} 
\end{table}

EBL preserves BLS's short input window, high accuracy, and rapid closed-form fine-tuning, while reducing resource usage and latency for edge deployment.


\section{Conclusion}\label{conclusion}
In this paper, we present EBL for distributed and adaptive harmonic estimation.
As a quantised FPGA implementation of a BLS-style edge harmonic estimator, EBL adapts to different harmonic estimation scenarios through efficient closed-form updates and, compared with conventional edge deployment schemes, achieves a 17.4$\times$ faster response with significantly lower resource consumption (LUTs and DSPs). In the future, we will extend this method to harmonic control strategy parameter prediction for real-time harmonic optimisation.


\bibliographystyle{IEEEtran}
\bibliography{references}

@article{li2023broad,
  title={Broad learning system using rectified adaptive moment estimation for harmonic detection and analysis},
  author={Li, Congcong and others},
  journal={IEEE Transactions on Industrial Electronics},
  volume={71},
  number={3},
  pages={2873--2882},
  year={2023},
  publisher={IEEE}
}

@article{genccol2023efficient,
  title={An efficient iterative optimization-based algorithm for the real-time estimation of harmonics under power system frequency deviations},
  author={Gen{\c{c}}ol, Kenan},
  journal={Engineering Science and Technology, an International Journal},
  volume={47},
  pages={101543},
  year={2023},
  publisher={Elsevier}
}

@inproceedings{li2024fast,
  title={A Fast Power Harmonic Detection Framework based on {FPGA}},
  author={Li, Jian and others},
  booktitle={2024 36th Chinese Control and Decision Conference (CCDC)},
  pages={1054--1059},
  year={2024},
  organization={IEEE}
}

@article{baraskar2022digital,
  title={Digital design of {DWPT} technique on {FPGA} for power system harmonics estimation},
  author={Baraskar, Savita and Tiwari, Vinay K},
  journal={Journal of Electrical Engineering \& Technology},
  volume={17},
  number={6},
  pages={3515--3524},
  year={2022},
  publisher={Springer}
}

@inproceedings{galus2008demand,
  title={Demand management of grid connected plug-in hybrid electric vehicles ({PHEV})},
  author={Galus, Matthias D and Andersson, Goran},
  booktitle={{IEEE Energy 2030 Conference}},
  pages={1--8},
  year={2008},
  organization={IEEE}
}

@article{davila2023harmonic,
  title={Harmonic distortion and hosting capacity in electrical distribution systems with high photovoltaic penetration: The impact of electric vehicles},
  author = {D{\'a}vila-Sacoto, Miguel and others},
  journal={Electronics},
  volume={12},
  number={11},
  pages={2415},
  year={2023},
  publisher={MDPI}
}

@article{iec200261000,
  title={61000-4-7: Electromagnetic compatibility ({EMC})},
  author={IEC, IEC61000},
  journal={Testing and measurement techniques-General guide on harmonics and interharmonics measurements and instrumentation, for power supply systems and equipment connected thereto, CEI-IEC, Geneva},
  year={2002}
}

@article{blott2018finn,
  title={{FINN-R}: An end-to-end deep-learning framework for fast exploration of quantized neural networks},
  author={Blott, Michaela and others},
  journal={ACM Transactions on Reconfigurable Technology and Systems (TRETS)},
  volume={11},
  number={3},
  pages={1--23},
  year={2018},
  publisher={ACM New York, NY, USA}
}

@article{norman2012hybrid,
  title={Hybrid wavelet and Hilbert transform with frequency-shifting decomposition for power quality analysis},
  author={Norman, CF and others},
  journal={IEEE Transactions on Instrumentation and Measurement},
  volume={61},
  number={12},
  pages={3225--3233},
  year={2012},
  publisher={IEEE}
}

@article{gu2007estimating,
  title={Estimating interharmonics by using sliding-window ESPRIT},
  author={Gu, Irene Yu-Hua and Bollen, Math HJ},
  journal={IEEE Transactions on Power Delivery},
  volume={23},
  number={1},
  pages={13--23},
  year={2007},
  publisher={IEEE}
}

@article{chang2009radial,
  title={Radial-basis-function-based neural network for harmonic detection},
  author={Chang, Gary W and others},
  journal={IEEE Transactions on Industrial Electronics},
  volume={57},
  number={6},
  pages={2171--2179},
  year={2009},
  publisher={IEEE}
}

@article{katyara2020signal,
  title={Signal parameter estimation and classification using mixed supervised and unsupervised machine learning approaches},
  author={Katyara, Sunny and others},
  journal={IEEE Access},
  volume={8},
  pages={92754--92764},
  year={2020},
  publisher={IEEE}
}

@article{madbhavi2023graph,
  title={Graph neural network-based distribution system state estimators},
  author={Madbhavi, Rahul and others},
  journal={IEEE Transactions on Industrial Informatics},
  volume={19},
  number={12},
  pages={11630--11639},
  year={2023},
  publisher={IEEE}
}

@article{tiwari2016hardware,
  title={Hardware implementation of polyphase-decomposition-based wavelet filters for power system harmonics estimation},
  author={Tiwari, Vinay K and Jain, Sachin K},
  journal={IEEE Transactions on Instrumentation and Measurement},
  volume={65},
  number={7},
  pages={1585--1595},
  year={2016},
  publisher={IEEE}
}

@article{tiwari2017fast,
  title={Fast amplitude estimation of harmonics using undecimated wavelet packet transform and its hardware implementation},
  author={Tiwari, Vinay K. and others},
  journal={IEEE Transactions on instrumentation and measurement},
  volume={67},
  number={1},
  pages={65--77},
  year={2017},
  publisher={IEEE}
}

@inproceedings{li2025logicsparse,
  title={{LogicSparse}: Enabling Engine-Free Unstructured Sparsity for Quantised Deep-learning Accelerators},
  author={Li, Changhong and Basu, Biswajit and Shanker, Shreejith},
  booktitle={2025 International Conference on Field Programmable Technology (ICFPT)},
  pages={223--224},
  year={2025},
  organization={IEEE}
}

@article{chen2018broad,
  title={Broad Learning System: An Effective and Efficient Incremental Learning System Without the Need for Deep Architecture},
  author={Chen, C. L. Philip and Liu, Zhe},
  journal={IEEE Transactions on Neural Networks and Learning Systems},
  volume={29},
  number={1},
  pages={10--24},
  year={2018},
  doi={10.1109/TNNLS.2017.2716952}
}

@inproceedings{naoussi2009fpga,
  title={{FPGA} implementation of harmonic detection methods using neural networks},
  author={Naoussi, S. R. D. and others},
  booktitle={13th European Conference on Power Electronics and Applications},
  pages={1--10},
  year={2009},
  address={Barcelona, Spain},
  doi={10.1109/EPE.2009.5279340}
}

@article{valtierra2013fpga,
  title={{FPGA}-based neural network harmonic estimation for continuous monitoring of the power line in industrial applications},
  author={Valtierra-Rodriguez, M. and others},
  journal={Electric Power Systems Research},
  volume={98},
  pages={51--57},
  year={2013},
  doi={10.1016/j.epsr.2013.01.011}
}

\end{document}